\documentclass[12pt,english]{paper}
\usepackage[T1]{fontenc}
\usepackage[latin1]{inputenc}
\usepackage{graphicx}

\makeatletter

\makeatletter

\makeatletter

\usepackage{geometry}

\makeatletter

\usepackage{epsfig}

\makeatother

\makeatother

\makeatother

\usepackage{babel}
\makeatother
\begin{document}

\title{Evaluation of the Neutron Background in a Direct WIMP Detector with Neutron Veto
System Based on Gd-doped Liquid Scintillator}

\author{Ye Xu$^1$%
\thanks{Corresponding author, e-mail address: xuye76@nankai.edu.cn%
}, XiangPan Ji$^1$, ZeLong Liu$^1$, DanNing Di$^1$, HaoKai Sun$^1$,
Yu Lei$^2$ }

\maketitle
\begin{flushleft}
$^1$School of Physics, Nankai University, Tianjin 300071, China
\par
$^2$University of Texas Health Science Center at San Antonio, San
Antonio, Texas 78229, USA

\par\end{flushleft}

\begin{abstract}
A direct WIMP (Weakly Interacting Massive Particle) detector with a
neutron veto system is designed to better reject neutrons. Two
experimental configurations are studied in the present paper: one is
for an Xe detector enclosed in a Gd-loaded scintillator and the
other one is for an Xe detector placed inside a reactor neutrino
detector. The Gd-doped liquid scintillator (or the neutrino
detector) is used as a neutron veto device. The neutron backgrounds
for the two experimental designs have been estimated using Geant4
simulations. The results show that the neutron backgrounds can
decrease to O(0.1) events per year per tonne of liquid Xenon. We
calculate the sensitivities to spin-independent WIMP-nucleon elastic
scattering. An exposure of one tonne $\times$ year could reach a
cross-section of about 6$\times$$10^{-11}$ pb.
\end{abstract}
\begin{flushleft}
\par\end{flushleft}

\section{Introduction}
It is found by the five year WMAP (the Wilkinson Microwave
Anisotropy Probe) data combined with the measurements of Type Ia
supernovas and baryon acoustic oscillations that the Universe is
made up of $4.6\%$ baryonic matter, $22.8\%$ dark matter and
$72.6\%$ dark energy\cite{GJR}. This means that about $83\%$ of the
matter in the Universe is unknown and called dark matter. Weakly
Interacting Massive Particles (WIMPs), predicted by extensions of
the Standard Model of particle physics, are a class of candidates
for dark matter\cite{GDJ}. They are distributed in the surrounding
halo of our galaxy. A WIMP halo of our galaxy with a local density
of $0.3 GeV/cm^3$ is assumed and its relative speed to the Sun is
$230 km/s$\cite{JP}.
\par
There is only weak interaction (and gravity) between WIMPs and
baryonic matter, so their direct detection is very challenging.
WIMPs can be directly detected through the measurement of nuclear
recoils induced by their elastic scattering off target nucleus.
Signals caused by interactions between WIMPs and nucleus can be
measured by ionization detectors, scintillation detectors or phonon
detectors. The background of this detection is made up of electron
recoils produced by $\gamma$ and $\beta$ particles and nuclear
recoils produced by neutrons. To reduce more background events due
to electron recoils, a hybrid detector has been employed in the
measurement of direct dark matter signatures. For example, these
signatures can be detected through ionization and scintillation in
the XENON10 experiment\cite{JEF} and also through ionization and
phonon in the CDMSII experiment\cite{cdms}. The electron recoil
contamination can decrease to less than $10^{-6}$ with hybrid
detectors\cite{cdms,PB,WHL}. However, it is very difficult to
discriminate between nuclear recoils induced by WIMPs and by
neutrons. This discrimination is one of the most important tasks in
direct dark matter searches.
\par
The cross-sections of neutron-nuclei interactions are much larger
than those of WIMP-nuclei, so the multi-interactions between
neutrons and detector components are applied to tag neutrons and
thus separate WIMPs from neutrons. In the ZEPLINIII experiment, the
0.5$\%$ Gd doped hydrocarbon material is used as the neutron veto
device. This device is used to tag neutrons, and the veto efficiency
is about 70$\%$-80$\%$\cite{MHJ,CBJ,HA}. The $2\%$ Gd-doped water is
also used as a neutron veto device with which the neutron background
can be reduced to 2.2 events per year per tonne of liquid Xenon
(LXe)\cite{AMA}.
\par
 Gd-doped liquid scintillator (Gd-LS) detectors can not
 only detect neutron-captured signals but also measure prompt
signals caused by neutrons, therefore the neutron tagging efficiency
should be higher than those of the Gd-doped hydrocarbon material
which only detect neutrons captured on Gd or H and Gd-doped water
which only detect neutrons captured on Gd. The background
environment for reactor neutrino experiments is similar to direct
dark matter searches, and in both cases the backgrounds  are
produced by cosmic muons and radioactivities in detector components
and their surrounding rock. Hence the background shield in reactor
neutrino experiments can be utilized for direct dark matter
searches. If a WIMP detector with a LXe target (Xe detector) is
placed inside the Gd-LS of a reactor neutrino detector, not only can
the shielding device of the neutrino detector shield background but
the Gd-LS of the neutrino detector can also be used to tag neutrons.
However, it seems that more neutron background events are produced
in the above experimental configuration:
\par
\begin{itemize}
\item The first additional background is from reactor neutrinos.
Neutrino detectors are fairly close to nuclear reactors (about 2
kilometers away) in reactor neutrino experiments. A large number of
reactor neutrinos will pass through the detectors, and nuclear
recoils will be produced by neutrino elastic scattering off target
nucleus in the Xe detectors. However, the energies of almost all of
these recoils are below 10 keV\cite{JMPF}. If the energy threshold
for WIMPs is set to more than 10 keV, the contamination from reactor
neutrinos will decrease to a negligible level.
\item The second additional
background is from low energy neutrons produced by the inverse
$\beta$-decay reaction $\bar{\nu_{e}}+p\rightarrow e^{+}+n$. But
their kinetic energies are almost below 100 keV\cite{chooz}, and
their maximum energy deposition in the Xe detectors is as large as a
few keV. Thus the neutron contamination can be reduced to a
negligible level by the energy threshold of more than 10 keV.
\item The third additional background is from cosmic muons.
The overburdens in reactor neutrino experiments are only a few
hundred meters water equivalent (m.w.e.). For example, the
overburden of the far hall in the Daya Bay neutrino experiment is
about 910 m.w.e.\cite{dayabay}. In general, the overburdens in
underground dark matter experiments are larger than 1500 m.w.e..
Consequently, more neutrons will be produced by cosmic muons in
reactor neutrino experiments in comparison with direct dark matter
searches. But the muon veto or thick water in reactor neutrino
experiments can tag or shield from these neutron events.
\end{itemize}
So reactor neutrino detectors can be used as neutron veto devices
and thus better reject neutron background in underground dark matter
experiments.
\par
We designed two experimental configurations in our work. First, four
Xe detectors are individually enclosed in four Gd-LS detectors which
are used as a neutron veto system (Gd-LS detector veto system). The
experimental hall of this experimental configuration is assumed to
be located in an underground laboratory with a depth of 2500 m.w.e.,
which is similar to the Canfranc underground laboratory\cite{JBJ}.
Second, four Xe detectors are individually placed inside four
reactor neutrino detectors which are used as a neutron veto system
(neutrino detector veto system). The experimental hall of the
corresponding experimental configuration is assumed to be located in
an underground laboratory with a depth of 910 m.w.e., which is
similar to the far hall in the Daya Bay experiment. The neutron
backgrounds for these two experimental designs are estimated using
Geant4\cite{geant} simulations.
\section{Detector description and some features of simulation}
Four Xe detectors are individually placed inside four Gd-LS
detectors (or the four reactor neutrino detectors) which are used as
a neutron veto system. Neutrons will be tagged through neutron
interactions with the Gd-LS. Two types of signals can be produced by
the interactions: a prompt signal induced by a proton recoil and a
delayed signal induced by a neutron captured on Gd or H. So the
Gd-LS can be used to shield from and tag neutrons using these
signals.
\subsection{Optimal Gd concentration}
Neutrons are captured on Gd with a capture time of about 30 $\mu$s
and on H with a capture time of about 180 $\mu$s\cite{dayabay}, so
the neutron-tagged efficiency by Gd is much higher than that by H.
In order to minimize the number of neutrons captured on H, the
optimal Gd concentration is studied using Geant4 simulations. Only
the neutron capture efficiency in the Gd-LS needs to be estimated,
whereas the behaviors of neutrons in other sub-detectors do not need
to be simulated. Accordingly a 2-meter-high cylinder of radius 1
meter filled with Gd-LS has been simulated, and neutrons with
energies up to 10 MeV have been shot from its center. We assume the
neutron energy follows a uniform distribution, since the ratio of
the neutron captures on Gd and H is neutron energy independent. The
results are summarized in Fig. 1. Saturating at a value of about
1$\%$, the neutron capture efficiencies do not change with the Gd
concentration, as seen in Fig. 1. Based on this simulation, the Gd
concentration in Gd-LS is set to 1$\%$ in this work.
\subsection{WIMP Detection with a Gd-LS detector veto system}
A detector with a Gd-LS veto system is designed to simulate a direct
detection experiment for WIMPs and estimate the neutron background.
The detector configuration is shown in Fig. 2. The detector is
located in a cavern of 40$\times$15$\times$12 $m^3$ at a depth of
2500 m.w.e.. The four identical cylindrical modules (each 477 cm
high and 429 cm in diameter) are immersed into a
13.6$\times$13.6$\times$9 $m^3$ water pool at a depth of 2.5 meters
from the top of the pool and at a distance of 2.5 meters from each
vertical surface of the pool. Each module is partitioned into three
enclosed zones. There is one Xe detector in an innermost zone (each
122 cm high and 74 cm in diameter), which is surrounded by the
middle zone filled with Gd-LS (1.3 m thick). The outermost zone is
filled with transparent mineral oil. 366 8-inch photomultiplier
tubes (PMTs) of 8-inch are mounted on the inside of the oil region
of the module. These PMTs are arranged in the same way as in
Ref.\cite{YWYKW}, that is, 8 rings of 30 PMTs are on the lateral
surface of the oil region, and 5 rings of 24, 18, 12, 6, 3 PMTs are
on the top and bottom caps. There is 0.25-tonne LXe used as an
active target in each Xe detector. Each Xe detector consists of
three components: LXe used as the active target (42 cm high and 51.6
cm in diameter), gaseous Xenon (16 cm high and 51.6 cm in diameter)
and liquid Nitrogen used as the cooling system (30 cm high and 52 cm
in diameter). The LXe and gaseous Xenon are contained in a copper
cylindrical vessel. 61 2-inch PMTs are mounted on the upper gaseous
Xenon and other 61 PMTs are mounted under the LXe target. The outer
stainless steel tank of each Xe detector is surrounded by an
Aluminum reflector for photons produced in the Gd-LS.

\subsection{WIMP Detection with a neutrino detector veto system}
A detector with a neutrino detector veto system is designed to
simulate a direct detection experiment for WIMPs and estimate the
neutron background. The detector configuration is shown in Fig. 2.
The detector is located in a cavern of 20$\times$20$\times$20 $m^3$
at a depth of 910 m.w.e.. The four cylindrical reactor neutrino
detector modules (414 cm high and 394 cm in diameter) are immersed
into a 13$\times$13$\times$8.5 $m^3$ water pool (water Cherenkov
detector) at a depth of 2.5 meters from the top of the pool and at a
distance of 2.5 meters from each vertical surface of the pool. Each
module is partitioned into three enclosed zones. The innermost zone
is filled with Gd-LS, which is surrounded by a zone filled with
unload liquid scintillator (LS). The outermost zone is filled with
transparent mineral oil\cite{dayabay}. 366 8-inch PMTs are arranged
in the same way as in Ref.\cite{YWYKW}. As discussed above in
Sec.2.1, the Gd concentration in the Gd-LS needs to be increased to
1$\%$ (it should be about 0.1$\%$ in reactor neutrino
experiments\cite{chooz,dayabay,DZ}). The four Xe detectors are
individually placed inside the four modules, and each Xe detector is
surrounded by the Gd-LS in the module. The Xe detectors have the
same configuration as those described in Sec.2.2.
\subsection{Some features of simulation}
The Geant4 (version 8.2) package has been used in our simulations.
The physics list in the simulations includes transportation
processes, decay processes, low energy processes, electromagnetic
interactions (multiple scattering processes, ionization processes,
scintillation processes, optical processes, cherenkov processes,
Bremsstrahlung processes, etc.) and hadronic interactions (lepton
nuclear processes, fission processes, elastic scattering processes,
inelastic scattering processes, capture processes, etc.). The cuts
for the productions of gammas, electrons and positrons are 1 mm, 100
$\mu$m and 100$\mu$m, respectively. The quenching factor is defined
as the ratio of the detector response to nuclear and electron
recoils. The Birks factor for protons in the Gd-LS is set to
0.01$g/cm^{2}/MeV$, corresponding to the quenching factor 0.17 at 1
MeV, in our simulations.
\section{Background estimation}
 The recoil energies for WIMP interactions with Xenon nuclei was set to
 a range from 15 keV to 50 keV\cite{AMA} in this work. Proton recoils induced
  by neutrons and neutron-captured signals are used to
  tag neutrons which reach the Gd-LS. The energy deposition produced by proton recoils is
  close to a uniform distribution. Neutrons captured on Gd and H lead to a release of
  about 8 MeV and 2.2 MeV of $\gamma$ particles, respectively. Due to the instrumental limitations of the
  Gd-LS, we assume neutrons will be tagged if their energy deposition in the Gd-LS is more than
1 MeV, corresponding to 0.17 MeVee (electron equivalent energy). In
the Gd-LS, it is difficult to distinguish signals induced by
neutrons from electron recoils, which are caused by the
radioactivities in the detector components and the surrounding
rocks. But these radioactivities can be controlled to less than 50
Hz according to the Daya Bay experiment\cite{dayabay}. If we assume
a 100 $\mu$s for neutron tagging time window, the indistinguishable
signals due to the radioactivities will result in a total dead time
of less than 44 hours per year.
\par
The contamination produced by
  neutrino events is reduced to a negligible level by the energy threshold of 15 keV\cite{JMPF}.
The electron recoil contamination mainly comes from $^{85}Kr$ in
commercially available Xenon gas (which decays through a beta-decay
with an endpoint energy of 678 keV), $^{238}U$, $^{232}Th$, $^{40}K$
in PMTs (which decay through gamma and beta decays), and $^{136}Xe$
in LXe (which undergoes a double-beta decay with a small
probability). In this study, only the electron recoil contamination
from Xenon gas and PMTs is considered. We then roughly
 estimate the electron recoil contamination. According to Bueno et al.\cite{AMA}, the rejection power
against electron recoils can reach $\sim$ 5$\times$$10^{-7}$ . If
the concentration of Kr in Xenon gas can be reduced to $\sim$1 ppb
and the radioactivity from PMTs is assumed to be O(100)
events/day\cite{EEA}, as a rough estimation, the total electron
recoil contamination is O(0.01) events/(ton$\cdot$yr). Compared to
the neutron background, this electron recoil contamination can be
ignored.
\par
Neutrons are produced from the detector components and their
surrounding rock. For the neutrons from the surrounding rock there
are two origins: first by spontaneous fission and ($\alpha$, n)
reactions due to U and Th in the rock (these neutrons can be omitted
because they are efficiently shielded, see Sec.3.2), and secondly by
cosmic muon interactions with the surrounding rock.
\par
We estimated the numbers of neutron background events in the LXe
target of one tonne. These numbers have been normalized to one year
of data taking and are summarized in Tab.1 and Tab.2.
\subsection{Neutron background from detector components}
 Neutrons from the detector components are induced by ($\alpha$, n)
 reactions due to U and Th. According to Mei et al.\cite{DCA}, the
 differential spectra of neutron yield can be expressed as
\begin{center}
 $\displaystyle Y_{i}(E_{n})=N_i{
\sum_{j}\frac{R_{\alpha}(E_{j})}{S_{i}^{m}(E_{j})}}$$\displaystyle
\intop_{0}^{E_{j}}\frac{d\sigma(E_{\alpha},E_{n})}{dE_{\alpha}}dE_{\alpha}$
\end{center}
where $N_i$ is the total number of atoms for the $i^{th}$ element in
the host material, $R_\alpha$$(E_j)$ refers to the $\alpha$-particle
production rate for the decay with the energy $E_j$ from $^{232}Th$
or $^{238}U$ decay chain, $E_{\alpha}$ refers to the $\alpha$
energy, $E_n$ refers to the neutron energy, and $S_{i}^{m}$ is the
mass stopping power of the $i^{th}$ element.
\subsubsection{Neutrons from PMTs in copper vessels}
The U and Th contaminations in the $SiO_2$ material are considered
as the only neutron source in the PMTs in our work. Neutrons from
$SiO_2$ are emitted with their average energy of 2.68 MeV\cite{DCA}.
The total number of the PMTs in the copper vessels of the four Xe
detectors is 488. The U and Th concentrations in the PMT components
can reach ten or even less ppb\cite{MJC}, so a rate of one neutron
emitted per PMT per year is conservatively estimated\cite{AMA}.
Consequently, there are 488 neutrons produced by all the PMTs in the
copper vessels per year.
\par
The simulation result in the case of
the Gd-LS detector veto system is summarized in Tab. 1. 13.8 neutron
events/(ton$\cdot$yr) reach the LXe target and their energy
deposition is in the same range as that of the WIMP interactions, as
seen in Tab. 1. Because 0.19 of them are not tagged in the Gd-LS,
these background events cannot be eliminated. The simulation result
in the case of the neutrino detector veto system summarized in Tab.
2. 14.2 neutron events/(ton$\cdot$yr) reach the LXe target and their
energy deposition is in the same range as that of the WIMP
interactions, as seen in Tab. 2. Because 0.22 of them are not tagged
in the Gd-LS, these background events cannot be eliminated. The
uncertainties of the neutron background from the PMTs in Tab.1 and
Tab.2 are from the binned neutron spectra in Ref.\cite{DCA}. But the
neutron background errors from the statistical fluctuation (their
relative errors are about 1$\%$) are too small to be taken into
account.
\subsubsection{Neutrons from copper vessels}
In the copper vessels, neutrons are produced by the U and Th
contaminations and emitted with their average energy of 0.81
MeV\cite{DCA}. Because each copper vessel is 5 mm thick, its total
volume is about 8000 $cm^3$. If we assume a 0.02 ppb U and Th
concentrations in the copper material, a rate of one neutron emitted
per 2000 $cm^3$ per year is conservatively estimated\cite{AMA}.
Consequently, there are 16 neutrons produced by the four copper
vessels per year.
\par
The simulation result in the case of the Gd-LS
detector veto system is listed in Tab. 1. 0.8 neutron
events/(ton$\cdot$yr) reach the LXe target and their energy
deposition falls in the same range as that of the WIMP interactions,
as seen in Tab. 1. As 0.02 of them are not tagged in the Gd-LS,
these background events cannot be eliminated. The simulation result
in the case of the neutrino detector veto system is summarized in
Table 2. 0.8 neutron events/(ton$\cdot$yr) reach the LXe target and
their energy deposition falls in the same range as that of the WIMP
interactions. As 0.02 of them are not tagged in the Gd-LS, these
background events cannot be eliminated. The uncertainties of the
neutron background from the copper vessels in Tab.1 and Tab.2 are
from the binned neutron spectra in the Ref.\cite{DCA}. But the
neutron background errors from the statistical fluctuation (their
relative errors are less than 1$\%$) are too small to be taken into
account.
\subsubsection{Neutrons from other components}
The U and Th contaminations in other detector components also
contribute to the neutron background in our experiment setup.
Neutrons from the Aluminum reflectors are emitted with the average
energy of 1.96 MeV\cite{DCA}. The U and Th contaminations in the
Carbon material are considered as the only neutron source in the
Gd-LS/LS. Neutrons from the Gd-LS/LS are emitted with the average
energy of 5.23 MeV\cite{DCA}. The U and Th contaminations in the Fe
material are considered as the only neutron source in the stainless
steel tanks. Neutrons from the stainless steel tanks are emitted
with the average energy of 1.55 MeV\cite{DCA}. We evaluated the
neutron background from the above components using Geant4
simulation. All the nuclear recoils in the LXe target, which is in
the same range as that of the WIMP interactions, are tagged. The
neutron background from these components can be ignored.
\subsection{Neutron background due to natural radioactivity in the rock}
Water can be used for shielding neutrons effectively, especially in
the low energy range of less than 10 MeV. \cite{JMC}. Additionally,
almost all neutrons due to natural radioactivity in the rock are
below 10 MeV\cite{AMA,MJC}. The Xe detectors are surrounded by about
2.5 meters of water and more than 1 meter of Gd-LS/LS, so these
shields can reduce the neutron contamination from the
radioactivities to a negligible level.
\subsection{Neutron background due to cosmic muons}
The total cosmogenic neutron fluxes can be evaluated as a function
of the depth for a site with a flat rock overburden\cite{DA}. The
energy spectrum and angular distribution of these neutrons can be
obtained by employing the method in Ref.\cite{DA,YFW} and the MUSUN
code\cite{VAK}. Fig. 3 shows the energy spectra of the cosmogenic
neutrons at a depth of 910 m.w.e. and 2500 m.w.e.. The neutrons are
sampled on the surfaces of the caverns, and the neutron interactions
with the detector are simulated with the Geant4 package.
\par
 The total cosmogenic neutron flux at a depth of 2500 m.w.e. (in the
case of the Gd-LS detector veto system) is
7.52$\times$$10^{-9}cm^{-2}s^{-1}$. Tab. 1 shows that 1.1 neutron
events/(ton$\cdot$yr) reach the LXe target and their energy
deposition is in the same range as that of the WIMP interactions.
But all of them are tagged by the Gd-LS. Because of that the neutron
background from cosmic muons can be ignored. The total cosmogenic
neutron flux at a depth of 910 m.w.e. (in the case of the neutrino
detector veto system) is 1.31$\times$$10^{-7}cm^{-2}s^{-1}$. Muon
veto systems can tag muons very effectively, thereby most cosmogenic
neutrons can be rejected. For example, using the muon veto system,
the neutron contamination level could be reduced by a factor of
about 10\cite{MJC}. In the Daya Bay experiment, the contamination
level can even be reduced by a factor of more than 30\cite{dayabay}.
Tab. 2 shows that 26 neutron events/(ton$\cdot$yr) reach the LXe
target and their energy deposition is in the same range as that of
the WIMP interactions. 0.4 of them are not tagged by the Gd-LS/LS.
We conservatively assume the neutron contamination level from cosmic
muons decreases by a factor of 10 using a muon veto system. This
could lead to the decrease of cosmogenic neutron contamination to
0.04 events/(ton$\cdot$yr). The uncertainties of the cosmogenic
neutron background in Tab.1 and Tab.2 are from the statistical
fluctuation.
\subsection{Total Background}
According to our simulations, if the detector is located in an
underground laboratory at a depth of 2500 m.w.e., the neutron
background is mainly from the PMTs and copper vessels, and the total
background is 0.21$\pm$0.03 events/(ton$\cdot$yr) (see Tab.1). If
the detector is located in an underground laboratory at a depth of
910 m.w.e., the neutron background is mainly from the PMTs, copper
vessels and cosmic muons, and the total background is 0.28$\pm$0.05
events/(ton$\cdot$yr) (see Tab.2).
\section{Conclusion and discussion}
The neutron background can be effectively suppressed with the Gd-LS
or neutrino detector used as the neutron veto system in direct dark
matter searches. Tab. 1 and Tab. 2 show the total neutron
contaminations in the cases of the Gd-LS veto system and neutrino
detector veto system are 0.2 and 0.3 events/(ton$\cdot$yr),
respectively. The neutron background events are caused by two
sources:
\begin{itemize}
\item After interacting with the LXe nucleus, a part of neutrons
are absorbed by the components within the Xe detector. Most of the
neutrons are absorbed by the liquid nitrogen in the cooling system,
and a few of the neutrons are absorbed by the copper vessels and
steel tanks in the Xe detector. These neutrons cannot deposit their
energy in the Gd-LS and thus are misidentified as WIMP signatures.
\item The other neutrons deposit a part of energy in the Gd-LS after
interacting with the LXe nucleus, but their energy deposition is
below the energy threshold of 1 MeV. Hence these neutrons are
misidentified as WIMP signatures.
\end{itemize}
\par
If no signals are significantly observed, sensitivities to
WIMP-nucleon spin-independent elastic scattering can be calculated
via the same method as Ref.\cite{GR}. To evaluate these
sensitivities, we assume a standard dark matter galactic
halo\cite{JP}, an energy resolution that amounts to 25$\%$ for the
energy range of interest and 50$\%$ nuclear recoil acceptance. Our
calculation shows that an exposure of one tonne $\times$ year could
reach a cross-section of about 6$\times$$10^{-11}pb$ at the 90$\%$
confidence level (see Fig.4). We note some updated results for dark
matter searches: the CDMSII experiment and XENON100 experiment have
given the upper limits on the WIMP-nucleon spin-independent
cross-section of 3.8$\times$$10^{-8}$pb for a WIMP of mass 70
GeV/$c^2$ and 7.0$\times$$10^{-9}$pb for a WIMP of mass 50 GeV/$c^2$
at the 90$\%$ confidence level, respectively\cite{cdms,xenon100}.
\par
Compared to Ref.\cite{AMA}, the neutron background in the present
paper decreases by a factor of about 10. Especially in the case of
the neutrino detector veto system, after finishing a precision
measurement of the neutrino mixing angle $\theta_{13}$, we can
utilize the existing experiment hall, background shield and veto,
Gd-LS and so on. This will not only save substantial cost and time
for direct dark matter searches, but the exclusion limit for the
experimental configuration could also reach about
6$\times$$10^{-11}pb$ at the 90$\%$ confidence level (one tonne
$\times$ year).
\section{Acknowledgements}

This work is supported in part by the National Natural Science
Foundation of China (NSFC) under the contract No. 10605014.

\begin{table}
\caption{Estimation of neutron background from different sources for
an underground laboratory at a depth of 2500 m.w.e. The column
labeled "15keV<$E_{recoil}$<50keV" identifies the number of neutrons
whose energy deposition in the LXe is in the same range as WIMP
interactions. The column labeled "Not Tagged" identifies the number
of neutrons which are misidentified as WIMP signatures (their energy
deposition in the LXe is in the same range as WIMP interactions
while their recoil energies in the Gd-LS are less than the energy
threshold of 1 MeV). The row labeled "PMTs in copper vessels"
identifies the number of neutrons from the PMTs in the copper
vessels. The row labeled "copper vessels" identifies the number of
neutrons from the copper vessels. The row labeled "cosmic muons"
identifies the number of cosmogenic neutrons. The terms after $\pm$
are errors. }
\begin{tabular}{|c|c|c|}\hline
   & 15keV<$E_{recoil}$<50keV & Not Tagged \\
  \hline
   PMTs in copper vessels & 13.8$\pm$2.21 & 0.19$\pm$0.030 \\
  \hline
   copper vessels & 0.8$\pm$0.15 & 0.02$\pm$0.003 \\
  \hline
   cosmic muons & 1.1$\pm$0.10 & $<$0.01 \\
  \hline
   total        & 15.7$\pm$2.2 & 0.21$\pm$0.03 \\
  \hline

\end{tabular}
\end{table}
\par
\begin{table}
\caption{Estimation of neutron background from different sources for
an underground laboratory at a depth of 910 m.w.e. The row labeled
"cosmic muons" identifies the number of cosmogenic neutrons in the
case of not using the muon veto system. The row labeled "muon veto"
identifies the number of cosmogenic neutrons in the case of using
the muon veto system. The other columns and rows in this table are
of the same meaning as in Tab. 1. We assume that neutron
contamination level from cosmic muons decreases by a factor of 10
using a muon veto system. Only the total background in the case of
using the muon veto system is listed in this table. The terms after
$\pm$ are errors.}
\begin{tabular}{|c|c|c|}\hline
   & 15keV<$E_{recoil}$<50keV & Not Tagged \\
  \hline
   PMTs in copper vessels & 14.2$\pm$2.72 & 0.22$\pm$0.035 \\
  \hline
   copper vessels & 0.8$\pm$0.15 & 0.02$\pm$0.003 \\
  \hline
   cosmic muons & 26.0$\pm$3.2 & 0.4$\pm$0.4 \\
  \hline
   muon veto & 2.6$\pm$0.32 & 0.04$\pm$0.04 \\
  \hline
   total (muon veto) & 17.6$\pm$2.7 & 0.28$\pm$0.05 \\
  \hline
\end{tabular}
\end{table}

\newpage

\begin{figure}
\begin{center}
\includegraphics[width=10cm,height=8cm]{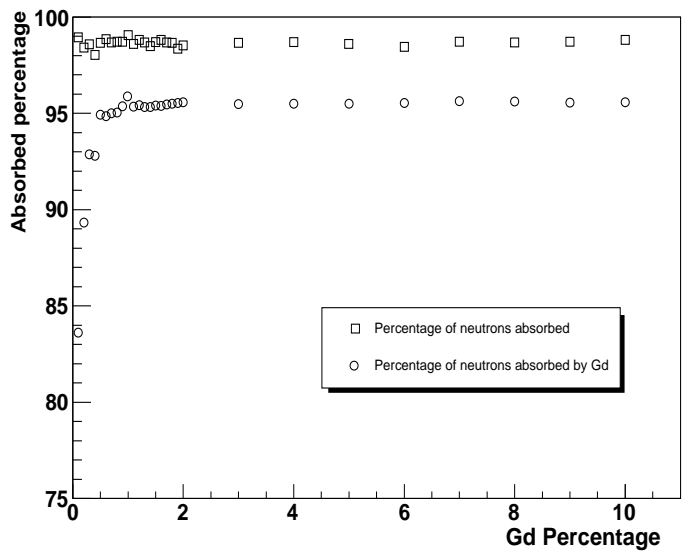}
\end{center}

\caption{Percentage of absorbed neutrons as a function of the Gd
concentration. The squares denote the total neutron-absorption
efficiencies of Gd-LS and the circles denote the neutron-absorption
efficiencies of Gd.}
\end{figure}

\par
\begin{figure}
\begin{center}
\includegraphics[width=8cm,height=10cm]{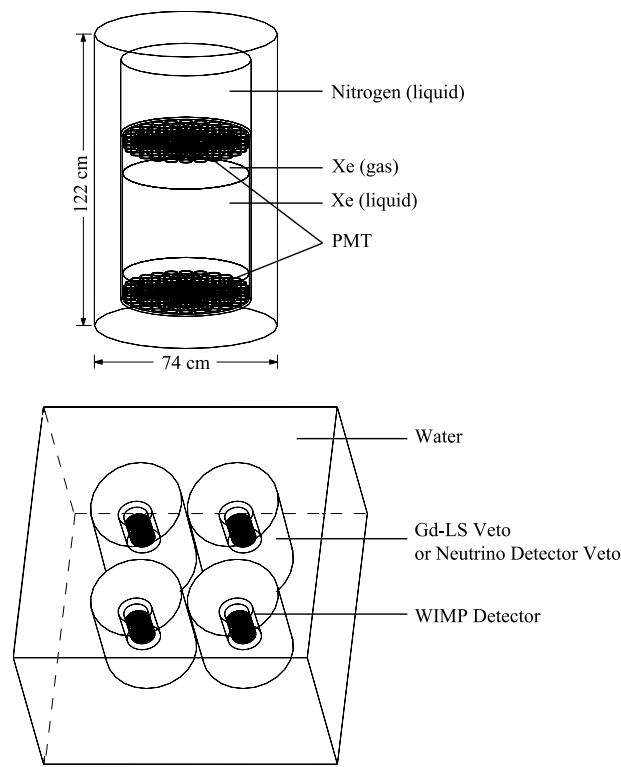}
\end{center}

\caption{Top: WIMP detector with a LXe detector (Xe detector).
Bottom: four Xe detectors individually placed inside four Gd-LS
detectors or four neutrino detectors in a water shield.}
\end{figure}

\newpage

\begin{figure}
\begin{center}
\includegraphics[width=10cm,height=8cm]{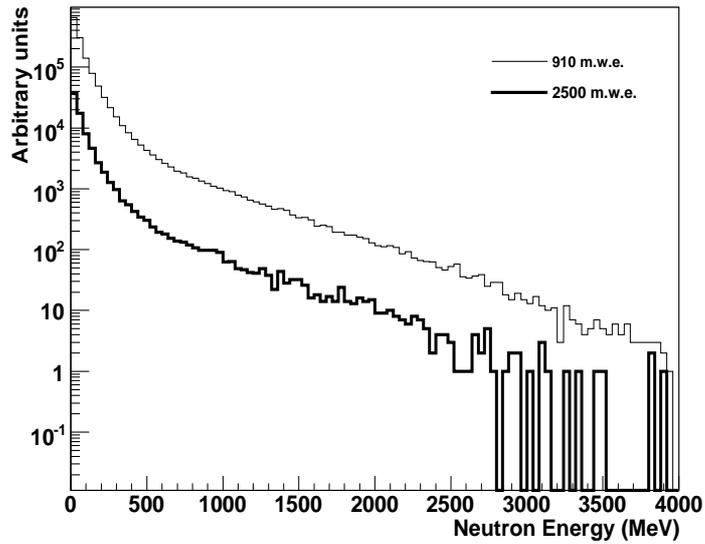}
\end{center}

\caption{The energy spectra of cosmogenic neutrons at depths of 910
m.w.e. and 2500 m.w.e.}
\end{figure}
\par
\begin{figure}
\includegraphics[width=10cm,height=8cm]{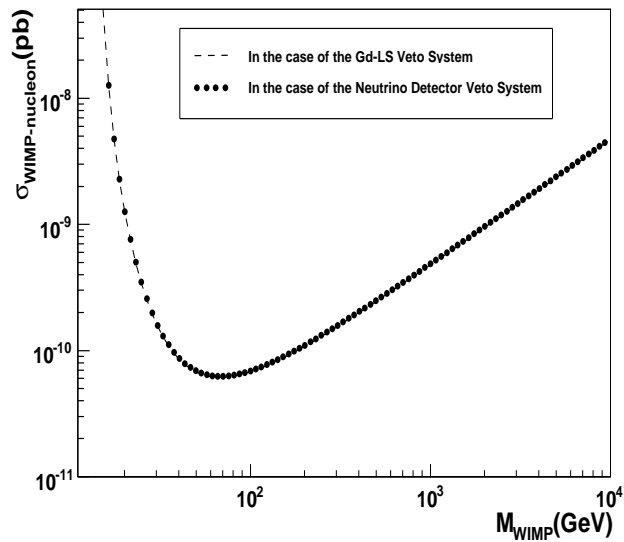}

\caption{We calculate the sensitivities to spin-independent
WIMP-nucleon elastic scattering assuming an exposure of one tonne
$\times$ year. The calculation shows this exposure could reach a
cross-section of about 6$\times$$10^{-11}pb$ at the 90$\%$
confidence level. The tool from Ref.\cite{limits} has been used. }
\end{figure}
\par

\end{document}